\documentclass[twocolumn]{revtex4-2}
\usepackage{graphicx}
\usepackage{dcolumn}
\usepackage{bm}
\usepackage{amsmath}
\usepackage{amsfonts}
\usepackage{amssymb}
\usepackage{color}
\usepackage{epstopdf}
\usepackage{float}
\providecommand{\U}[1]{\protect\rule{.1in}{.1in}}

\begin{document}
\title{Entropy-based analysis of single-qubit Otto and Carnot heat engines}
\author{Andr\'es Vallejo}
\author{Catty Lissardy}
\author{Santiago Silva-Gallo}
\author{Alejandro Romanelli}
\author{Ra\'ul Donangelo}
\affiliation{\begin{small} Facultad de Ingenier\'{\i}a, 
Universidad de la Rep\'ublica, Montevideo, Uruguay\end{small}}
\date{\today}

\begin{abstract}

From an entropy-based formulation of the first law of thermodynamics in the quantum regime, we investigate the performance of Otto-like and Carnot-like engines for a single-qubit working medium. Within this framework, the first law includes an additional contribution —coherence work— that quantifies the energetic cost of deviating the quantum trajectory from its natural unitary evolution. We focus on the efficiency of the heat-to-coherence work conversion and show that the Carnot cycle achieves the classical Carnot efficiency, while the performance of the Otto cycle is upper-bounded by the Carnot efficiency corresponding to the extreme temperatures of the cycle. We identify entropy generation during the isochoric stages as the key source of irreversibility limiting the Otto cycle’s efficiency.
\end{abstract}

\maketitle

\section{\label{sec:level1}Introduction}

Entropy-based quantum thermodynamics \cite{Ahmadi2023, Alipour2022} offers an alternative framework for analyzing thermodynamic processes in the quantum regime. This approach starts by assuming that von Neumann entropy provides a valid extension of thermodynamic entropy for non-equilibrium quantum processes, and subsequently defines heat as the energy exchange associated with an entropy variation. This leads to alternative definitions of heat and work, which can yield conclusions that differ significantly from those of standard thermodynamic analyses \cite{Vallejo2020,Vallejo2021}. The entropy-based approach has already been applied in connection with various quantum thermodynamic scenarios, for instance, to quantify the degree of non-Markovianity in open system's dynamics \cite{Choquehuanca2023}.

A distinctive feature of this approach is the emergence of an additional trajectory-dependent term in the first law of thermodynamics, known as environment-induced work or coherence work, which captures the energetic contribution of quantum coherence \cite{deLima2021}. In two-level systems, this quantity can be interpreted as the energy required to rotate the Bloch vector away from its natural unitary trajectory defined by the local Hamiltonian, due to environmental interactions. In more general systems, coherence work is related to the expected value of the commutator between the local Hamiltonian and an effective Hamiltonian governing the evolution of the eigenstates of the density matrix \cite{Vallejo2025}. Since this expected value quantifies the degree of non-commutativity between the two generators, it serves as a measure of how much the actual trajectory deviates from the unitary one, implying that the larger the mismatch, the greater the coherence work involved. The connection between coherence work and ergotropy has been examined in Ref.~\cite{Choquehuanca2024}.

In this work, we explore the entropy-based formalism by analyzing the performance of Otto-like and Carnot-like power cycles. These cycles have been extensively studied using a wide variety of working media, including a particle in a potential well \cite{Bender2000,Abe2011}, quantum harmonic oscillators \cite{He2009, Kosloff2017,Dann2020}, spin systems \cite{Geusic1967,Geva1991,Heinrich2007,Quan2007,Cakmak2017,Camati2019,Altintas2019,Cakmak2020,Cakmak2021,Chakraborty2022,Koyanagi2022,Picatoste2024}, and quantum planar rotors \cite{Gaida2024}. Moreover, experimental realizations of these cycles have been achieved —for instance, using spin-1/2 systems controlled via nuclear magnetic resonance techniques— yielding, in some cases, work extraction efficiencies approaching the maximum theoretical limit \cite{Peterson2019}. In this manuscript, the working medium will be taken as a single qubit.

These previous analyses rely on the traditional approach, where work is defined as the portion of the energy change due to explicit variations in the system's Hamiltonian, while heat is associated with changes in the system's state, i.e., in its reduced density matrix \cite{Alicki}. 
In contrast, the entropy-based framework employed here defines heat strictly as the energy exchange linked to entropy variation, with the remaining contribution corresponding to the standard mechanical work due to Hamiltonian changes, and coherence work.
A related formulation, developed in Ref. \cite{Dann2020_2}, derives heat and work directly from the Heisenberg equation of motion: the dissipative part, which entails entropy change, is associated with heat, while the remaining unitary part is identified as work. Unlike the entropy-based approach, this formulation does not treat coherence as an explicit energetic contribution, restricting the definition of work to energy changes induced by Hamiltonian control.

In the present analysis, we restrict our attention to situations in which the local Hamiltonian remains fixed throughout the cycle and focus exclusively on studying the efficiency of the heat-to-coherence work conversion. This efficiency quantifies  the fraction of the heat absorbed during the cycle that is converted into work on the environment associated with driving the working medium along a prescribed quantum trajectory. 

Since coherence work arises from environment-induced deviations of the system’s state rather than from changes in a control parameter, its nature is fundamentally different from standard mechanical work. It cannot be accumulated at will in a work reservoir but is instead transferred to the environment that produces the deviation. Nevertheless, it is meaningful to study the efficiency of converting heat into this contribution because, within this formalism, it constitutes a new term in the first law. In this study, we examine whether it shares properties with conventional work, in particular regarding efficiency, while leaving aside the question of practical plausibility. Indeed, although qubit-based engines provide a useful theoretical setting to explore fundamental thermodynamic concepts, the energetic cost of the precise control operations typically required exceeds the work produced. For example, a detailed study in Ref.~\cite{Pedram2023} shows that the energy required to implement control protocols can outweigh the extracted work in a quantum Otto engine based on a harmonic oscillator, leading to negative efficiency values.

The structure of the paper is as follows. In Section II, we provide a brief review of the entropy-based formulation of the first law of thermodynamics, presenting alternative notions of heat, work, and temperature for two-level systems. In Section III, we introduce the Otto and Carnot cycles and conduct their thermodynamic analysis, discussing the results. Remarks and conclusions are presented in Section IV.

\section{Entropy-based quantum thermodynamics in the Bloch sphere: a review}
For completeness, we provide a brief review of the theoretical framework from which the thermodynamic analyses are performed. The results discussed in this section have been previously reported in Ref. \cite{Vallejo2021}.

Consider an open qubit interacting with an environment. Its local Hamiltonian can be expressed in terms of the Pauli matrices $\vec{\sigma}=(\sigma_x,\sigma_y,\sigma_z)$, as:

    \begin{equation}\label{H}
    H=-\vec{v} \cdot \vec{\sigma},
    \end{equation}
where $\vec{v}$ can be interpreted as an effective magnetic field, already expressed in energy units.
Similarly, the qubit's reduced density matrix can be written in terms of the Bloch vector $\vec{B}$, as:
    \begin{equation}\label{rho}
    \rho=\frac{1}{2}(\mathcal{I}+\vec{B} \cdot \vec{\sigma}).
    \end{equation}
From Eqs. (\ref{H}) and (\ref{rho}), it can be shown that the internal energy of the system, defined as the expected value of the local Hamiltonian, adopts the form:
    \begin{equation}\label{energy}
    E=\text{tr}(\rho H)=-\vec{B} \cdot \vec{v}.
    \end{equation}

Entropy-based quantum thermodynamics relies on two basic assumptions: i) that the von Neumann entropy serves as an appropriate extension of the thermodynamic entropy in out-of-equilibrium quantum processes; ii) that heat corresponds to the portion of the energy change that is concomitant with the entropy variation. For a two-level system, the von Neumann entropy depends solely on the modulus of the Bloch vector,  $B=\vert\vec{B}\vert$.  Its expression is:
    \begin{equation} \label{entropy}
    S(B) = -k_B \left[
    \frac{1+B}{2} \ln\!\left(\frac{1+B}{2}\right)
    + \frac{1-B}{2} \ln\!\left(\frac{1-B}{2}\right)
    \right].
    \end{equation}
In light of this dependence, it is convenient to write the energy, Eq. (\ref{energy}), as $E=-B\hat{B} \cdot\vec{v}$, where $\hat{B}$ is the unit vector in the direction of $\vec{B}$. Differentiating this expression yields:
    \begin{equation}\label{dE}
    dE =-dB(\hat{B}\cdot \vec{v})-B(d\hat{B})\cdot\vec{v}-\vec{B} \cdot d\vec{v}.
    \end{equation}

Based on the previous discussion, and noting that only the first term in Eq. (\ref{dE}) is associated with the entropy variation, we define the heat exchanged by the system as:
	\begin{equation}\label{heat}
	\dot{\mathcal{Q}}=-\dfrac{dB}{dt}(\hat{B}\cdot\vec{v}).
	\end{equation}

As a consequence, the remaining two terms can then be identified as work. The third term:
    \begin{equation}\label{mechanical work}
	\dot{W}=-\vec{B}\cdot \dfrac{d\vec{v}}{dt}
	\end{equation}
represents the standard mechanical work due to Hamiltonian control, while the second term:		
	\begin{equation}\label{coherence work}
	\dot{C}=-B\dfrac{d\hat{B}}{dt}\cdot\vec{v},
	\end{equation}
referred to as environment-induced work or coherence work, has been shown to be related to the presence of coherence during the process. In this study, this quantity is of central importance, as we focus on the generation of coherence work from heat.
	
Finally, we introduce the concept of effective temperature. By analogy with the classical case, we define the temperature as the derivative of energy with respect to entropy in a zero work process. Rewriting Eq.~\eqref{dE} as $dE = -(\hat{B} \cdot \vec{v})\, dB - B\, d(\hat{B} \cdot \vec{v}),$ it becomes clear that no work is done when the quantity $\hat{B} \cdot \vec{v}$ remains constant. Thus, the temperature is given by:
    \begin{equation}\label{temperature1}
    \dfrac{1}{T}=\dfrac{\partial S}{\partial E}\biggr\vert_{\hat{B}\cdot \vec{v}}.
    \end{equation}
Applying the chain rule, Eq. (\ref{temperature1}) can be rewritten as 
    \begin{equation}\label{temperature2}
    \frac{1}{T} = \frac{dS}{dB} \cdot \left(\frac{dB}{dE}\right)_{\hat{B} \cdot \vec{v}}.    
    \end{equation}
The required derivatives can be computed from Eqs.~(\ref{energy}) and (\ref{entropy}):
    \begin{equation}\label{dSdB}
    \frac{dS}{dB} = -k_B \tanh^{-1}(B), 
    \quad
    \left(\frac{\partial B}{\partial E}\right)_{\hat{B}\cdot \vec{v}} = -\frac{1}{\varepsilon \cos\theta}.
    \end{equation}
Substituting these expressions into Eq.~\eqref{temperature2} yields the final expression for the qubit's temperature \cite{Vallejo2021}:  
    \begin{equation}\label{temperature}
    T=\dfrac{\varepsilon \cos\theta}{k_B\tanh^{-1}(B)},
    \end{equation}
where $\theta$ is the angle between $\vec{B}$ and $\vec{v}$, and $\varepsilon =\vert\vec{v}\vert$.

It is important to remark that the present framework does not rely on a microcanonical description of either the working medium or the global system. Instead, the relevant ensemble is the set of accessible reduced density operators $\rho$; accordingly, the von Neumann entropy constitutes an appropriate entropy functional at this level of description. The temperature introduced here is an effective temperature, defined as the integrating factor relating entropy and energy variations along zero-work processes. It can be shown that this temperature coincides with the canonical (Boltzmann) temperature for thermal equilibrium states; however, unlike the latter, it remains well defined for arbitrary non-equilibrium states. Consequently, it should not be interpreted as a microcanonical absolute temperature, whose definition relies on the fact that, for isolated systems, the Gibbs volume entropy provides the only thermodynamically consistent definition of entropy \cite{Hanggi2016, Hilbert2014}. Within this framework, the effective temperature may formally take negative values for population-inverted states of finite systems with bounded spectra, without introducing any fundamental inconsistency. In the Bloch-sphere representation employed here, this situation arises whenever the angle $\theta$ between the Bloch vector
$\vec B$ and the effective field $\vec v$ satisfies $\cos\theta<0$,
so that the system is predominantly aligned with the higher-energy eigenstate of the local Hamiltonian. Note, however, that the Otto-like and Carnot-like cycles analyzed in this work operate entirely within the positive-temperature branch, so the appearance of negative values is neither required nor invoked in the physical implementation of the cycles. Despite representing a concept distinct from the standard notion of temperature, this effective temperature nevertheless plays a key role in characterizing the efficiencies of the cycles discussed below.

\section{Results}
\subsection{Otto cycle}

Classically, an Otto cycle consists of two isentropic and two isochoric processes. Since entropy depends solely on $B$, it follows that isentropic trajectories are confined to spheres concentric with the Bloch sphere.

On the other hand, isochoric processes are characterized by the absence of work. As we are interested in analyzing the conversion of heat into coherence work, we will focus exclusively on cases where the mechanical work, as defined by Eq. (\ref{mechanical work}), is zero. This condition is satisfied by keeping
$\vec{v}$ constant. Consequently, isochoric processes are defined by the vanishing of Eq. (\ref{coherence work}), where the only non-trivial, non-isentropic solution occurs when 
$\hat{B}$ remains constant. In conclusion, isochoric trajectories, in the sense of not including coherence work, are represented by straight lines passing through the origin of the Bloch sphere.

\begin{figure}[h!]
        \centering
        {\includegraphics[width=0.75\columnwidth]{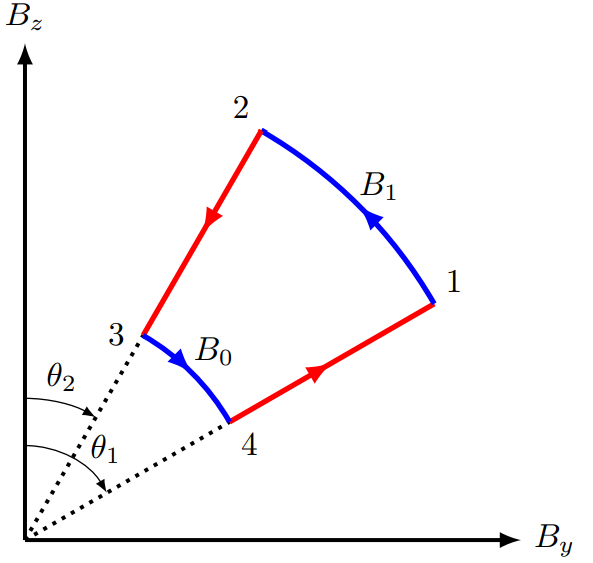}}
        \caption{Representation of the Otto cycle on the Bloch sphere. The cycle is represented in the plane $B_x=0$ for convenience, though any vertical plane could have been chosen. The four strokes are: $1 \rightarrow 2$: isentropic process with $ |\vec{B}| = B_1$, during which the Bloch vector rotates from angle $\theta_1$ to $\theta_2$; $2 \rightarrow 3$: isochoric heating at fixed angle $\theta_2$; $3 \rightarrow 4$: isentropic process with $ |\vec{B}| = B_0$, where the vector rotates back from $\theta_2$ to $\theta_1$; and $4 \rightarrow 1$: isochoric cooling at fixed angle $\theta_1$.} 
        \label{f1}
    \end{figure}

In light of the aforementioned considerations, the simplest implementation of an Otto cycle defines the region illustrated in Fig. (\ref{f1}).

The first stage, $1 \rightarrow 2$, corresponds to an \textit{isentropic process}, in which the modulus of the Bloch vector remains constant at $B_1$, while the angle $\theta$ decreases from $\theta_1$ to $\theta_2$.
Since the entropy depends only on $B$, it remains unchanged. 
This process can be implemented through a purely unitary dynamics by introducing a control Hamiltonian that makes the effective rotation generator $\Omega$ proportional to $\sigma_x$, thus inducing a coherent rotation around the $x$-axis.

The second stage, $2 \rightarrow 3$, is an \textit{isochoric heating} process in which the angle remains fixed at $\theta_2$, while $B$ decreases from $B_1$ to $B_0$. This radial trajectory can be generated by a purely dissipative Lindblad dynamics (i.e., with no unitary part) defined by a single jump operator proportional to $\sigma_x$. This dynamics preserves the direction of the Bloch vector (as long as it lies in the $yz$-plane) and produces an exponential contraction of its modulus, generating a straight-line path toward the center of the Bloch sphere. By applying this dynamics for a finite time, the system can be driven to any intermediate point along this path.

In the third stage, $3 \rightarrow 4$, the system undergoes an \textit{isentropic process}, with constant modulus $B_0$ and the angle returning from $\theta_2$ to $\theta_1$. The entropy remains constant, and the process can again be implemented by unitary dynamics. As before, the effective rotation generator can be written as $\Omega =  \kappa\sigma_x$, but in this case the sign of $\kappa$ must be negative in order to reverse the sense of rotation.

Finally, the fourth stage, $4 \rightarrow 1$, is an \textit{isochoric cooling} process in which the angle is held fixed at $\theta_1$, and the modulus increases from $B_0$ back to $B_1$. The entropy and temperature decrease during this stage. It can be shown that this outward trajectory requires two jump operators of the form
    \begin{equation}
    L_0 = \sqrt{\lambda}\, |\psi\rangle \langle 0|, \qquad L_1 = \sqrt{\lambda}\, |\psi\rangle \langle 1|,
    \label{operators_pure_state}
    \end{equation}
where $|\psi\rangle$ is a pure state pointing in the direction of the Bloch vector, and $\lambda > 0$. This dissipative dynamics asymptotically purifies the system towards the state $|\psi\rangle$, but by applying it for a finite time, it is possible to reach any intermediate mixed state along the straight segment from the center to the target.

An alternative method can be constructed by using the spectral decomposition of the target density operator. In this method, jump operators are defined as 
\begin{equation}
\begin{aligned}
L_0 &= \sqrt{\lambda p_1}\, |\psi_1\rangle \langle 0|, \quad
L_1 = \sqrt{\lambda p_1}\, |\psi_1\rangle \langle 1|, \\
L_2 &= \sqrt{\lambda p_2}\, |\psi_2\rangle \langle 0|, \quad
L_3 = \sqrt{\lambda p_2}\, |\psi_2\rangle \langle 1|.
\end{aligned}
\end{equation}
where $\{|\psi_1\rangle, |\psi_2\rangle\}$ are the eigenstates of the target density matrix, $p_1$ and $p_2$ are the corresponding eigenvalues, and $\lambda > 0$ is a positive parameter controlling the dissipation rate. In this case, it can be shown that the system asymptotically reaches the target state, possibly not requiring zero-temperature environments to produce the evolution.

Writing $\hat{B}=\cos\theta\hat{k}+\sin\theta\hat{i}$, $\vec{v}=\varepsilon\hat{k}$, and applying Eqs. (\ref{heat}) and (\ref{coherence work}) to the isentropic trajectories, we obtain:
    \begin{equation}\label{isent}
	\begin{cases}
	\mathcal{Q}_{isent}=0\\
	C_{isent}=-\varepsilon B(\cos\theta_f- \cos\theta_i),
	\end{cases}
    \end{equation}
where $i$ and $f$ refer to the initial and final states of the trajectory. \\
For the isochoric trajectories, the corresponding quantities are
    \begin{equation}\label{isoch}
	\begin{cases}
	\mathcal{Q}_{isoc}= -\varepsilon\cos\theta(B_f-B_i)\\
	C_{isoc}=0,
	\end{cases}
    \end{equation}
Accordingly, the non-zero energy exchanges along the cycle of Fig. (\ref{f1}) are:
    \begin{equation}
	\begin{cases}
	C_{1\rightarrow 2}=\varepsilon B_1(\cos\theta_1-\cos\theta_2)\\[3pt]
	\mathcal{Q}_{2\rightarrow 3}=\varepsilon\cos\theta_2(B_1-B_0)
	\\[3pt]
	C_{3\rightarrow 4}=\varepsilon B_0(\cos\theta_2-\cos\theta_1)\\[3pt]
	\mathcal{Q}_{4\rightarrow 1}=\varepsilon\cos\theta_1(B_0-B_1)
	\end{cases}
    \end{equation}

In order to obtain the efficiency, we need the heat absorbed and the coherence work produced. 
Heat enters the system only along the stage $2\rightarrow 3$, so we have: 
    \begin{equation}
    \mathcal{Q}_{H}=\mathcal{Q}_{2\rightarrow 3}=\varepsilon\cos\theta_2(B_1-B_0).
    \end{equation}
On the other hand, the global coherence work can be obtained as 	
    \begin{equation}
    \vert C_{net}\vert =\vert C_{1\rightarrow 2}+C_{3\rightarrow 4}\vert= \varepsilon(B_1-B_0)(\cos\theta_2-\cos\theta_1).
    \end{equation} 
Finally, the efficiency of the Otto cycle adopts the form:
    \begin{equation}\label{etaOtto}
    \eta_{Otto}=\dfrac{\vert C_{net}\vert}{Q_{H}}=1-\dfrac{\cos\theta_1}{\cos\theta_2}.
    \end{equation}
Note that, according to Eq. (\ref{etaOtto}), the efficiency of the Otto cycle is independent of the radius of the isentropic trajectories and is determined exclusively by the angles defining the isochoric processes. This is similar to what occurs in standard analyses of this cycle, where the only work present is due to Hamiltonian driving. In those cases, efficiencies typically take the form \( \eta = 1 - \text{(ratio)} \), where the ratio depends on the control parameters that define the Hamiltonian during the isochoric processes \cite{Geva1991,Kosloff2017}.

In order to properly assess the performance of the Otto cycle within this framework, it is essential to first understand how the Carnot cycle behaves under the same entropy-based formulation. Since Carnot’s efficiency sets the fundamental upper bound for any heat engine, analyzing its implementation in this context provides a critical reference for interpreting the results obtained for other cycles. The next subsection is therefore devoted to examining the quantum Carnot cycle, and to verifying whether its classical efficiency is preserved in the present framework.

\subsection{Carnot Cycle}

\begin{figure}[h!]
        \centering
        {\includegraphics[width=0.85\columnwidth]{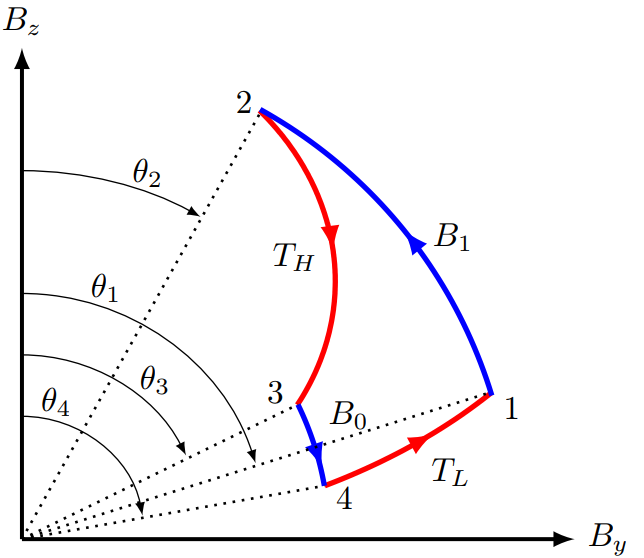}}
        \caption{Diagrammatic representation of the Carnot cycle on the Bloch sphere. The four strokes are: $1\rightarrow2$: isentropic process with $ |\vec{B}| = B_1$, where the Bloch vector rotates from $\theta_1$ to $\theta_2$; $2\rightarrow3$: isothermal heating at temperature $T_H$, from $\theta_2$ to $\theta_3$; $3\rightarrow4$: isentropic process with  $ |\vec{B}| = B_0$, from $\theta_3$ to $\theta_4$; and $4\rightarrow1$: isothermal cooling at temperature $T_L$, returning from $\theta_4$ to $\theta_1$.}
        \label{f2}
    \end{figure}

As in the classical case, the quantum Carnot cycle consists of two isentropic and two isothermal stages. The isentropic processes are analogous to those of the previous cycle, involving coherent rotations at constant Bloch vector modulus. The isothermal processes correspond to stages $2 \rightarrow 3$ and $4 \rightarrow 1$, and take place on surfaces of constant temperature. Since temperature is given by Eq.~(\ref{temperature}), these isothermal surfaces are defined by the expression
    \begin{equation}
    B(\theta) = \tanh\left(\dfrac{\varepsilon\cos\theta}{k_B T}\right),
    \end{equation}
with $T = T_H$ or $T = T_L$. A typical cycle is shown in Fig.(\ref{f2}).

For the isentropic stages, the heat and coherence work are given by Eq. (\ref{isent}), while for the isothermal trajectories we obtain:
    \begin{equation}\label{isot}
        \begin{cases}
         \mathcal{Q}_{isot}=-k_BT[f(B_f)-f(B_i)]\\[8pt]
         C_{isot}=\dfrac{k_BT}{2}\ln\left(\dfrac{1-B_f^2}{1-B_i^2}\right),
        \end{cases}
    \end{equation}
where $f(B)=B\tanh^{-1}B+\frac{1}{2}\ln(1-B^2)$.
From Eqs. (\ref{isent}) and (\ref{isot}), we found that
    \begin{equation}
        \begin{cases}
	    C_{1\rightarrow 2}=\varepsilon
        B_1(\cos\theta_1-\cos\theta_2)\\[3pt]
	    \mathcal{Q}_{2\rightarrow 3}=k_BT_H[f(B_1)-f(B_0)]\\[4pt]
        C_{2\rightarrow 3}=\frac{k_BT_H}{2}\ln\left(\frac{1-B_0^2}{1-      B_1^2}\right)\\[3pt]
	    C_{3\rightarrow 4}=\varepsilon B_0(\cos\theta_3-\cos\theta_4)\\[3pt]
	    \mathcal{Q}_{4\rightarrow 1}=k_BT_L[f(B_0)-f(B_1)]\\[4pt]
	    C_{4\rightarrow 1}=\frac{k_BT_L}{2}\ln\left(\frac{1-B_1^2}{1-B_0^2}\right).
	    \end{cases}
    \end{equation}
The global coherence work is obtained by adding the four contributions. After some calculus, we obtain:	
    \begin{equation}
    \vert C_{net}\vert=k_B(T_H-T_L)[f(B_1)-f(B_0)].
    \end{equation}
Meanwhile, the heat absorbed from the reservoir is
    \begin{equation}
    \mathcal{Q}_{net}=\mathcal{Q}_{2\rightarrow 3}=k_BT_H[f(B_1)-f(B_0)],
    \end{equation}
so for the efficiency we obtain the standard result:
    \begin{equation}
    \eta_{Carnot}=\dfrac{\vert C_{net}\vert}{Q_{net}}=1-\dfrac{T_L}{T_H}.
    \end{equation}

\subsection{Discussion}
It is a well-known fact that the efficiency of the classical Otto cycle is always lower than the Carnot efficiency corresponding to the extreme temperatures of the cycle. In order to analyze what occurs in this alternative approach, it is necessary to determine the limit temperatures of the cycle, which, we assume, are the temperatures of the reservoirs with which the qubit is in contact along the isochoric processes. According to Eq. (\ref{temperature}), it is possible to see that the lowest and highest temperatures involved occur at states 1 and 3, respectively:
    \begin{equation}
        \begin{cases}
        T_{L}=T_1=\dfrac{\varepsilon \cos\theta_1
        }{k_B\tanh^{-1}(B_1)}\\[10pt]
        T_{H}=T_3=\dfrac{\varepsilon \cos\theta_2}{k_B\tanh^{-1}(B_0)}
        \end{cases}
    \end{equation} 
From these results, the Carnot efficiency adopts the form:
    \begin{equation}\label{etaCarnot}
    \eta_{Carnot}=1-\dfrac{T_L}{T_H}=1-\alpha\dfrac{\cos\theta_1}{\cos\theta_2}
    \end{equation}
where $\alpha=\tanh^{-1} (B_0) /\tanh^{-1} (B_1)$. Since $\alpha\leq 1$, we conclude that
    \begin{equation}
    \eta_{Carnot}\geq\eta_{Otto},
    \end{equation}
consistently with the classical case. From Eqs. (\ref{etaOtto}) and (\ref{etaCarnot}), it can be seen that the relation between both efficiencies is:
    \begin{equation}
    \eta_{Carnot}=1-\alpha(1-\eta_{Otto}).
    \end{equation}
Since the limit temperatures of the cycle depend on $B$, the same occurs with the corresponding Carnot efficiencies. However, as $B_1\rightarrow B_0$, they approach the Otto efficiency, which represents their lower bound.

\begin{figure}[h!]
        \centering
        {\includegraphics[width=0.6\columnwidth]{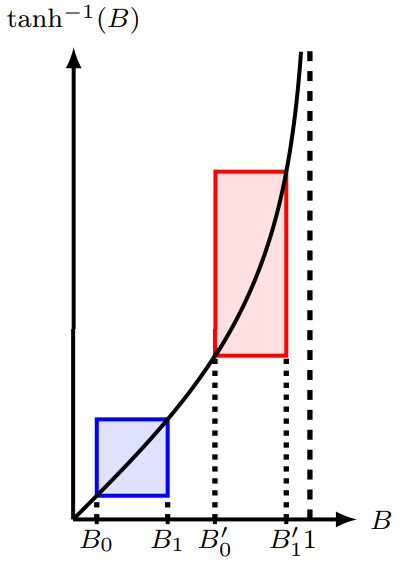}}
        \caption{The entropy production in the quantum Otto cycle (in units of $k_{B}$) can be represented by the area of a rectangle in the $\tanh^{-1}(B)$ vs. $B$ diagram. For cycles with fixed width $\Delta B$, the rectangles have larger areas as their bases shift closer to 1, indicating greater entropy production.}
        \label{f3}
    \end{figure}

The impossibility of attaining Carnot efficiency can be explained noting that the isochoric stages are, from the classical point of view, irreversible processes. This is so because the system's temperature differs from the reservoir temperature along the isochoric processes. The global irreversibility can be quantified in terms of the entropy production, which, in this case, can be found by adding the entropy changes of the reservoirs. The result is:
    \begin{equation}\label{Sgen}
     S_{gen}=k_B(B_1-B_0)(\tanh^{-1}(B_1)-\tanh^{-1}(B_0))
    \end{equation}
which is clearly non-negative. This expression admits a simple geometric interpretation, as illustrated in Fig. (\ref{f3}). The entropy produced (in units of $k_{\text{B}}$) corresponds to the area of the rectangle defined by the horizontal interval $(B_0, B_1)$ and the vertical interval $(\tanh^{-1}(B_0), \tanh^{-1}(B_1))$, in the plot of $\tanh^{-1}(B)$ versus $B$. For a fixed width $\Delta B = B_1 - B_0$, the generated entropy increases as the interval is shifted toward $B = 1$, reflecting the larger temperature difference between the system and the reservoir during the isochoric stages, and thus a higher degree of irreversibility.

In terms of the entropy production, the relation between the Otto and Carnot efficiencies is the usual in power cycles operating between two reservoirs \cite{Marcella}:
    \begin{equation}
    \eta_{Otto}=\eta_{Carnot}-\dfrac{T_LS_{gen}}{\mathcal{Q}_{H}}.
    \end{equation}
Again, we can see that in the limit $B_1\rightarrow B_0$ no entropy is produced, so both efficiencies coincide.

\section{Remarks and conclusions}

In this manuscript we have explored the entropy-based formulation of thermodynamics in the quantum regime by analyzing two simple power cycles. 

The analysis of the Carnot cycle confirmed that, under this formulation, the efficiency matches the classical Carnot limit. This was expected, as Carnot efficiency emerges from two key assumptions: the separation of energy exchanges into two contributions of different nature (heat and work), and the validity of the Clausius relation between entropy and heat for reversible processes. Since these properties are preserved in the entropy-based approach, the classical result is recovered independently of the specific entropy function used \cite{Tobin}. Moreover, the inclusion of mechanical work alongside coherence work would not alter this conclusion, as the derivation depends solely on the classification of energy flows and the validity of the Clausius relation, not on the specific form of the work term.

In the case of the Otto cycle, we found that the efficiency depends only on the angles associated with the isochoric processes and not on the radius of the Bloch vector trajectories. The efficiency is always lower than the Carnot efficiency corresponding to the extreme temperatures involved, and the deviation can be traced to entropy production during the isochoric stages.

Comparisons with previous work are not straightforward, as most analyses of power cycles consider work resulting from changes in the system's Hamiltonian, rather than coherence work, a recently introduced concept. Nevertheless, this work generalizes certain results from previous analyses.
For instance, Eq. (\ref{Sgen}) naturally extends the entropy production derived in pioneering studies of the quantum Otto cycle \cite{Geva1991}. In that case, the component $B_z$ played the role of $B$, as all processes were confined to the z axis.

A key assumption of this framework is that von Neumann entropy is a valid extension of the thermodynamic entropy in non-equilibrium quantum systems, a perspective that remains the subject of debate \cite{Hemmo2006, Prunkl2020, Chua2021}. 

Interestingly, if pressure is defined by analogy with the classical case, it turns out to be the modulus of the Bloch vector. This implies that isentropic processes are also isobaric, effectively restricting the class of meaningful cycles in this framework to those composed of isothermal, isentropic, and isochoric processes—namely, Carnot, Otto, and Stirling cycles. 

Future work will explore the Stirling cycle in detail, as well as the open-system control strategies required to ensure that the system’s state effectively follows the desired evolutions. Unlike most of the studies cited, where the system evolves through thermal states of the Hamiltonian and the thermodynamic strokes are implemented by varying either the temperature or a control parameter, the cycles analyzed here are fully dynamical: the system's state follows a continuous evolution along each segment, driven either by unitary control or by engineered dissipative dynamics.
While we have described the Lindblad and unitary operators associated with isochoric and isentropic segments, further analysis is required to develop viable experimental protocols to implement such evolutions in practice, particularly for the isothermal trajectories, which involve more complex paths in the Bloch sphere. However, the analysis performed guarantees that, when those strategies are implemented, the cycles have the efficiencies reported.

\section*{Acknowledgments}
This work was partially supported by the Uruguayan agencies Agencia Nacional de Investigaci\'on e Innovaci\'on (ANII) and 
Programa de Desarrollo de las Ciencias B\'asicas (PEDECIBA).

\end{document}